\documentclass[11pt]{article}
\usepackage[colorlinks,citecolor=blue,urlcolor=blue]{hyperref}

\def\siv{_{\text{IV}}}
\def\spp{_{\text{PP}}}
\def\sat{_{\text{AT}}}
\def\stsls{_{\text{TSLS}}}
\def\sps{_{\text{PS}}}
\def\saps{_{\text{APS}}}

\def\sraw{_{s-\text{raw}}}
\def\sshrunk{_{s-\text{shrunk}}}
\def\ssplit{_{s-\text{split}}}
\def\thraw{\thetahat\sraw}
\def\thshr{\thetahat\sshrunk}
\def\thspl{\thetahat\ssplit}

\usepackage[utf8]{inputenc}
\usepackage[english]{babel}
 
\usepackage[margin=1in]{geometry}
\usepackage{lmodern}
\usepackage{amsmath}               
\usepackage{graphicx}

\usepackage[T1]{fontenc}
\RequirePackage{amsthm,amsmath,geometry,amsfonts}
\RequirePackage{natbib}
\usepackage[utf8]{inputenc}
\usepackage[english]{babel}
\usepackage{centernot}

\def\bb{{\bf b}}

\def\bd{{\bf d}}

\def\bw{{\bf w}}

\def\bC{{\bf C}}

\def\bJ{{\bf J}}

\def\bP{{\bf P}}

\def\bS{{\bf S}}
\def\bT{{\bf T}}

\def\bV{{\bf V}}

\def\bX{{\bf X}}

\def\thick#1{\hbox{\rlap{$#1$}\kern0.25pt\rlap{$#1$}\kern0.25pt$#1$}}

\def\bbeta{\boldsymbol{\beta}}
\def\bgamma{\boldsymbol{\gamma}}

\def\btheta{\boldsymbol{\theta}}

\def\bDelta{\boldsymbol{\Delta}}

\def\bSigma{\boldsymbol{\Sigma}}

\def\smbalpha{\boldsymbol{{\scriptstyle{\alpha}}}}

\def\dhat{{\widehat d}}
\def\ehat{{\widehat e}}

\def\Ehat{{\widehat E}}

\def\Phat{{\widehat P}}

\def\Shat{{\widehat S}}
\def\That{{\widehat T}}

\def\Vhat{{\widehat V}}

\def\bdhat{{\widehat \bd}}

\def\bChat{{\widehat \bC}}

\def\bPhat{{\widehat \bP}}

\def\bThat{{\widehat \bT}}

\def\bVhat{{\widehat \bV}}

\def\thetahat{{\widehat\theta}}

\def\lambdahat{{\widehat\lambda}}

\def\pihat{{\widehat\pi}}
\def\rhohat{{\widehat\rho}}

\def\Sigmahat{{\widehat\Sigma}}

\def\bbetahat{{\widehat\bbeta}}

\def\bthetahat{{\widehat\btheta}}

\def\bDeltahat{{\widehat\bDelta}}

\def\bSigmahat{{\widehat\bSigma}}

\def\smbalpha{\widehat{\smbalpha}}

\def\hbar{\bar{ h}}

\def\Asc{{\cal A}}
\def\Bsc{{\cal B}}

\def\Ssc{{\cal S}}

\def\transpose{{\sf \scriptscriptstyle{T}}}

\def\E{\mbox{E}}

\def\var{\mbox{var}}

\def\sumin{\sum_{i=1}^n}

\def\trans{^{\transpose}}
\def\inv{^{-1}}

\def\argmindum{\mathop{\mbox{argmin}}}
\def\argmin#1{\argmindum_{#1}}

\def\logit{\mbox{logit}}

\def\var{\mbox{var}}

\def\mybox#1{\vskip1mm \begin{center}
        \hspace{.0\textwidth}\vbox{\hrule\hbox{\vrule\kern6pt
\parbox{.9\textwidth}{\kern6pt#1\vskip6pt}\kern6pt\vrule}\hrule}
        \end{center} \vskip-5mm}
\def\lboxit#1{\vbox{\hrule\hbox{\vrule\kern6pt
      \vbox{\kern6pt#1\vskip6pt}\kern6pt\vrule}\hrule}}

\def\thickboxit#1{\vbox{{\hrule height 1mm}\hbox{{\vrule width 1mm}\kern6pt
          \vbox{\kern6pt#1\kern6pt}\kern6pt{\vrule width 1mm}}
               {\hrule height 1mm}}}

\def\fat#1{\hbox{\rlap{$#1$}\kern0.25pt\rlap{$#1$}\kern0.25pt$#1$}}

\newtheorem{assumption}{Assumption}

\usepackage{titling}

\title{Synthetic estimation for the complier average causal effect}
\author{Denis Agniel$^*$, Bing Han, Matthew Cefalu\\
RAND Corporation\\
1776 Main St\\
Santa Monica, CA, 90401\\
$^*$dagniel@rand.org}
\date{}
\begin{document}
\begin{titlingpage}
    \maketitle
    \begin{abstract}
    We propose an improved estimator of the complier average causal effect (CACE). Researchers typically choose a presumably-unbiased estimator for the CACE in studies with noncompliance, when many other lower-variance estimators may be available. We propose a synthetic estimator that combines information across all available estimators, leveraging the efficiency in lower-variance estimators while maintaining low bias. Our approach minimizes an estimate of the mean squared error of all convex combinations of the candidate estimators. We derive the asymptotic distribution of the synthetic estimator and demonstrate its good performance in simulation, displaying a robustness to inclusion of even high-bias estimators.
\end{abstract}
 
\end{titlingpage}

\section{Introduction}
Noncompliance is common in studies where some study subjects may self-select into a different study condition than the one to which they were assigned. Noncompliance may be related to unmeasured factors, so without further assumptions, the presence of noncompliance can complicate the analysis of even a randomized experiment.  A popular approach to circumvent the noncompliance issue is the intention-to-treat (ITT) analysis \citep{roland1998understanding, hollis1999meant, heckman2001policy}.  An ITT analysis aims to estimate a ``real-world'' or ``diluted'' effect of a treatment, ignoring the noncompliance in the study sample and assuming that the level of noncompliance in the study sample reflects the actual situation if the treatment were to be implemented elsewhere \citep{ten2008intent}.  The estimand in an ITT analysis is usually referred to as the \textit{effectiveness} of a treatment.  
An ITT analysis is straightforward in the presence of noncompliance, where the (often random) assignments are used as the factor of interest, while the actual treatment status subject to noncompliance is ignored.  In randomized studies, since the randomized assignment is orthogonal to all confounders, a simple two-sample comparison can be unbiased for estimating the effectiveness of a treatment \citep{angrist1999empirical, little2009comparison}. 

In contrast, the \textit{efficacy} of a treatment refers to the effectiveness of a treatment when it is actually taken. One measure of efficacy is the \textit{complier average causal effect} (CACE), or the mean causal effect among those who will comply with treatment assignment, i.e., the \textit{principal compliers} \citep{angrist1996identification, little2009comparison}.  A favored approach to estimate the efficacy of a treatment is a structural equation model, where the assignment is considered as an instrumental variable (IV) for the actual treatment status received \citep{greene2003econometric}.  Under certain assumptions, the IV approach is unbiased for estimating the efficacy of a treatment.   See \citep{imbens2014instrumental} for a detailed survey of methods and development in this space. Despite its popularity, the IV approach can suffer from  a high estimation variance, particularly if the sample proportion of principal compliers is low \citep{little2009comparison, antonelli2017synthetic}.
 	
Two different approaches, the per-protocol (PP) and as-treated (AT) analyses, have also received many applications in assessing the efficacy of a treatment \citep{higgins2008cochrane, mcnamee2009intention}. The PP analysis subsets the sample to those whose actual treatment status is the same as their randomized assignment, i.e., \textit{observed compliers}.  However,  observed compliers are usually different from principal compliers. For example, those who would always refuse a treatment, i.e., \textit{never-takers}, cannot be differentiated from principal compliers when assigned to the control condition.  The difference between the observed and principal compliers results in biased estimation of the CACE in the PP analysis. By contrast, the AT analysis ignores the initial assignment and uses the actual treatment to estimate a treatment effect.  Due to various sample selection biases, the AT analysis can also be biased for the CACE.  Despite their biases, both the PP and AT analyses usually have a smaller estimation variance than the IV approach.  If we measure the efficiency of an estimator by its mean squared error (MSE), i.e., the sum of its squared bias and sampling variance, among the IV, PP, and AT estimators, none can always outperform the others.  See, e.g., \citet{little2009comparison} and \citet{antonelli2017synthetic} for discussions of the scenarios where each estimator can outperform the others.  The suitable scenario for each estimator depends on non-estimable properties of some unobserved data, e.g., whether there is a mean difference in the outcome between never-takers and principal compliers when both are assigned to the control condition.  

In this paper, we consider a synthetic estimator as a convex combination of a set of candidate estimators of the CACE, including but not limited to the IV, AT, and PP estimators.  
Our approach is rooted in the theory of model averaging, which conventionally focuses on estimating coefficients in least square regression models \citep{buckland1997model, hjort2003frequentist, judge2004semiparametric, mittelhammer2005combining, longford2006missing, hansen2007least}, and predicting random effects in mixed-effect models \citep{robinson1991blup, ghosh1994small, Searle1997, longford2006missing}. The theoretical framework of model averaging is much wider than linear regression and mixed-effect models.  In the recent literature, \citet{lavancier2016general} outlined synthetic estimators for a very general class of problems and provided some asymptotic results. \citet{antonelli2017synthetic} proposed a synthetic estimator in the presence of noncompliance in the setting of a randomized controlled trial. In the same spirit, we use the term synthetic estimation instead of model averaging because our candidate estimators do not belong to a common class of models, although they target the same estimand of interest, i.e., the CACE. One purpose of synthetic estimation is to combine several available candidate estimators to form a single and unambiguous estimator.  By striking a balance between biases and variances, a combination of candidate estimators can have a smaller MSE than some or even all candidate estimators, in particular, any \textit{a priori} favored candidate estimator.   

We propose a class of synthetic compliance estimators (SCEs) which aim to optimally combine candidate estimators of the CACE by minimizing the estimated MSE of the resulting estimator. The proposed SCE is appropriate for observational studies, randomized trials, and other experimental and non-experimental studies where estimation must be adjusted for covariates.  The SCE displays a robustness property: without sacrificing too much on the estimation bias, it borrows information from other biased but precise candidate estimators to improve the MSE. 

The rest of the paper is organized as follows. In Section 2, we outline specifics about the principal compliance framework and lay out the details of the candidate estimators in this setting. In Section 3, we present the SCE and its practical implementation. The asymptotic properties of the SCE are discussed in Section 4.  Section 5 includes simulations demonstrating the operating characteristics and robustness of the SCE. Section 6 contains concluding remarks.

\section{Framework}
\subsection{Principal stratification and CACE}
Suppose interest lies in the causal effect of an assigned treatment on an outcome $Y_i$, for $n$ subjects indexed by $i$. The actual treatment status, denoted by $S_i$, is a random variable which may not equal the assignment $Z_i$ possibly due to subject self-selection into a different treatment status. Further, suppose $\bX_i$ is a set of covariates collected on these $n$ subjects. We assume throughout that $Z_i$ and $S_i$ are binary, so that they take values in $\{0, 1\}$. Let $S_i(z)$ be the potential treatment status that would have been observed if $Z_i = z, z = 0, 1$. Based on the configuration of $Z_i$ and $S_i(z)$, there are potentially four types of individuals in the population: \textit{always-takers} who always take the treatment regardless of assignment, i.e. $S_i(z)=1$; \textit{never-takers} who always take the non-treatment condition regardless of assignment, i.e. $S_i(z)=0$ ; \textit{principal compliers} who always conform to the assignment, i.e. $S_i(z)=z$; \textit{principal defiers} who always do the opposite of the assignment, i.e. $S_i(z)=1-z$. These four groups of individuals are known as the \textit{principal strata}. The actual treatment status $S_i$ is not influenced by the assignment for always-takers and never-takers. By contrast, principal compliers' and principal defiers' actual treatment status is determined by their assignment.  Under the principal stratification framework, the group of \textit{observed compliers} consists of principal compliers, never-takers assigned to the control condition, and always-takers assigned to the treatment condition. In this paper we focus on the principal compliers.  Hereafter, the principal compliers are referred to as the "compliers" for brevity in presentation.  Let the indicator of compliance be $C_i = I\{S_i(z) = z, z = 0, 1\}$, and let $\pi_c = P\{S_i(z) = z, z = 0, 1\}$ be the corresponding probability. Further, let $n_z = \sumin I\{Z_i = z\}$ be the sample size in the group assigned to treatment $z$, and let $n_{zs} = \sumin I\{Z_i = z, S_i = s\}$ be the sample size of the group assigned to treatment $z$ and with treatment status $s$.

Define the potential outcome $Y_i(z, s)$ to be the outcome that one would observe if, possibly contrary to fact, $Z_i = z$ and $S_i = s$. For these counterfactuals to make sense, we assume the no-interference or Stable Unit Treatment Value Assumption (SUTVA) holds \citep{rubin1978bayesian} for both the potential outcomes $Y_i(z,s)$ and the potential treatments $S_i(z)$. That is, that the potential outcomes and potential treatment values for individual $i$ do not depend on any other individual, and $S_i = S_i(1)Z_i + S_i(0)(1-Z_i)$ and 
\begin{align*}
    Y_i = Y_i(1, 1)Z_iS_i + Y_i(0, 1)(1-Z_i)S_i + Y_i(1, 0)Z_i(1-S_i) + Y_i(0, 0)(1-Z_i)(1-S_i).
\end{align*} 
Further, let the counterfactual based only on the treatment status be
\begin{align*}
    Y_i(s) = Y_i(1, s)Z_i + Y_i(0, s)(1-Z_i), s = 0, 1
\end{align*} 
By definition, a causal effect is the difference between a pair of distinct potential outcomes, where one or more causal factors differ. The \textit{efficacy} of a treatment or the \textit{complier average causal effect} (CACE) refers to the mean causal effect of treatment among compliers \citep{angrist1996identification, little2009comparison}. Using the notation above and suppressing the subscripts, it is
\begin{equation}
    \text{CACE} = \theta = E[Y(1) - Y(0) | S(z) = z].
\end{equation}

\subsection{Identifying assumptions}
In this section, we outline a series of assumptions that are typically used to identify the CACE. Some estimators may require many of the ensuing assumptions, while some may require only a few. 
\begin{assumption}[Pseudo-randomization] \label{iv_randomized}
    \begin{align*}
    \left.Z_i \amalg \left\{S_i(0), S_i(1), Y_i(0, 0), Y_i(0, 1), Y_i(1, 0), Y_i(1, 1)\right\}\right|\bX_i.    
    \end{align*}
\end{assumption}
Assumption \ref{iv_randomized} states that treatment assignment is, if not randomized, as good as randomized \citep{imbens2014instrumental} within strata defined by the covariates $\bX_i$. This assumption is similar in spirit to the no-unmeasured-confounders assumption commonly invoked for treatment effect estimation.

Next, we will make two monotonicity assumptions. 
\begin{assumption}[Monotonicity]\label{monotonicity}
\begin{align*}
S_i(0) \leq S_i(1). 
\end{align*}
\end{assumption}
The first is required for identifiability and states that there are no defiers in the population and the treatment assignment does not make anyone less likely to take treatment. 
\begin{assumption}[Strong monotonicity]\label{strong_monotonicity}
\begin{align*}
S_i(0) = 0. 
\end{align*}
\end{assumption}
This second, stronger version of monotonicity ensures that there are no always-takers ($S_i(z) = 1, z = 0,1$) and only serves to simplify notation. While strong monotonicity may not be a reasonable assumption in all cases, any additional complexity arising due to its violation (by inclusion of always-takers) could easily be incorporated in the framework we lay out here. 

\begin{assumption}[Exclusion restriction]\label{er}
\begin{align*}
    Y_i(s) = Y_i(0, s) = Y_i(1, s), s = 0, 1 
    \end{align*}
\end{assumption} 
Assumption \ref{er} encodes the so-called exclusion restriction (ER), which states that there is no direct effect of the treatment assignment on the outcome in the population. In situations where treatment assignment is determined by double-blind randomization, the ER is very plausible. In other cases, it must be justified based on substantive expertise. Under the ER, we can equivalently write the CACE as $E[Y_i(1, 1) - Y_i(0,0) | S_i(z) = z]$.

We further might require an assumption on the effect of compliance:
\begin{assumption}[No compliance effect]\label{nce}
\begin{align*}
    E[Y_i(z, s) | S_i(0) = 0, S_i(1) = 1, \bX_i] = E[Y_i(z, s) | S_i(0) = z, S_i(1) = z, \bX_i], z = s = 0, 1. 
    \end{align*}
\end{assumption}
Assumption \ref{nce} is not needed for traditional IV  or two-stage estimators and states that compliers ($S_i(z) = z$) have the same outcomes as certain non-compliers and thus there is \textit{no compliance effect} (NCE). When strong monotonicity holds, we only require the no compliance effect to hold when $z = s = 0$. A version of this assumption was termed General Principal Ignorability in Section 6.1 of \cite{Ding2017}.

\subsection{Candidate CACE estimators}
Our approach relies on a collection of candidate estimators, and it leverages the information in all of the candidates to produce an efficient estimator with low bias. While there have been many estimators of the CACE proposed previously, for clarity and ease of presentation, we restrict ourselves to a series of well-known ones. The SCE could easily be adapted to include more, fewer, or other candidate estimators with ease. 

\paragraph{IV estimator.} The standard IV estimator inflates the ITT estimator (which is attenuated due to noncompliance) by the inverse of the compliance probability. Then the IV estimator can be written
\begin{align}
    \thetahat\siv = \frac{n_1\inv\sum_{i: Z_i = 1}Y_i - n_0\inv\sum_{i: Z_i = 0}Y_i}{\pihat_c}. \label{iv_est}
\end{align}
Under Assumptions \ref{iv_randomized}, \ref{er}, and \ref{monotonicity}, $\thetahat\siv$ is consistent for the CACE, and under Assumption \ref{strong_monotonicity}, $\pihat_c = \Phat\{S_i(z) = z, z = 0, 1\}$ may be estimated by $\frac{n_{11}}{n_1}$.

\paragraph{Two-stage least squares.} The two-stage least squares (TSLS) estimator of \cite{Angrist1995} fits a least squares model for the outcome and an additional model for treatment status. The CACE is estimated as $\thetahat_{TSLS} = \thetahat$ from the model
\begin{align}
    Y_i = \beta_0 + \theta \Shat_i + \bgamma\trans\bX_i + \epsilon_i, i = 1, ..., n
\end{align}
and $\Shat_i$ is the predicted value of a regression of $S_i$ onto $Z_i, \bX_i$. Traditionally, the model for $S_i$ is a least squares model, but in practice one could use logistic regression or any other binary regression model. Under Assumptions \ref{iv_randomized}, \ref{er}, and \ref{monotonicity} and assuming the models for $Y_i$ and $S_i$ are correct, $\thetahat\stsls$ is consistent for the CACE.

\paragraph{Per-protocol.} The \textit{per-protocol} estimator models the effect of receiving the treatment among those individuals whose treatment and instrument agree. The PP estimator is found as $\thetahat\spp = \thetahat$ from the model:
\begin{align}
    Y_i = \beta_0 + \theta S_i + \bgamma\trans\bX_i + \epsilon_i, i \in \{j : S_j = Z_j\}.\label{pp-model}
\end{align}
Under Assumptions \ref{iv_randomized}, \ref{nce}, and \ref{monotonicity} and assuming the model \eqref{pp-model} holds, $\thetahat\spp$ is consistent for the CACE. 

\paragraph{As-treated.} The \textit{as-treated} estimator ignores the instrument and simply estimates the effect of $S_i$ from a model. It is found as $\thetahat\sat = \thetahat$ from the model:
\begin{align}
    Y_i = \beta_0 + \theta S_i + \bgamma\trans\bX_i + \epsilon_i, i = 1, ..., n \label{at-model}
\end{align}
Under Assumptions \ref{iv_randomized}, \ref{er}, \ref{nce}, and \ref{monotonicity} and assuming the model \eqref{at-model} holds, $\thetahat\sat$ is consistent for the CACE.

\paragraph{Principal-score weighting estimators.} Weighting methods using so-called principal scores are laid out in \cite{Ding2017}. The principal score for compliers or compliance score is $e(\bX) = P(C_i = 1 | \bX),$ the conditional probability of being a complier. We consider two principal-score estimators. The first is purely a weighting estimator and can be written:
\begin{align*}
    \thetahat\sps = n_{11}\inv\sum_{i : S_i = 1, Z_i = 1} Y_i - n_0\inv\sum_{i : Z_i = 0} \frac{Y_i\ehat(\bX_i)}{\pihat_c}.
\end{align*}
Under Assumptions \ref{iv_randomized}, \ref{nce}, \ref{strong_monotonicity}, and consistency of $\ehat(\cdot)$, $\thetahat\sps$ is consistent for the CACE. 

The second estimator is a model-assisted version of the same estimator:
\begin{align*}
    \thetahat\saps &= n_{11}\inv\sum_{i : S_i = 1, Z_i = 1} (Y_i - \bbetahat_1\trans\bX_i) - n_{0}\inv\sum_{i : Z_i = 0} \frac{(Y_i - \bbetahat_0\trans\bX_i)\ehat(\bX_i)}{\pihat_c} +\\
    &\qquad (n_0 + n_{11})\inv\sum_{i : Z_i = 0 \text{ or } S_i = 1, Z_i = 1} \frac{(\bbetahat_1 - \bbetahat_0)\trans\bX_i\ehat(\bX_i)}{\pihat_c}
\end{align*}
where $n_{zs}$ is the sample size in group $Z_i = z, S_i = s$ and $\bbetahat_1$ is estimated from the model
\[
Y_i = \beta_0 + \bbeta_1\trans\bX_i + \epsilon_i, i \in \{j : S_j = 1, Z_j = 1\}
\]
and $\bbetahat_0$ is estimated from the model
\[
Y_i = \beta_0 + \bbeta_1\trans\bX_i + \epsilon_i, i \in \{j : Z_j = 0\}.
\]
Under Assumptions \ref{iv_randomized}, \ref{nce}, \ref{strong_monotonicity}, and consistency of $\ehat(\cdot)$ -- no additional assumptions from $\thetahat\sps$ -- $\thetahat\saps$ is consistent for the CACE and has a lower asymptotic variance.

\paragraph{Principal-score stratified estimators.} We finally estimated versions of the IV, AT, and PP estimators that were stratified by the principal score $e(\bX)$. We first computed each estimator withiin quintiles of the principal score and then averaged across quintiles. 

\subsection{Bias-variance tradeoff in candidate estimators}

The estimators in the previous section present bias-variance trade-offs for the analyst. To illustrate this point, consider the properties of a few of the estimators under the model
\begin{align}
    Y_i = \beta_0 + \alpha_c C_i + \theta S_i + \bgamma\trans\bX_i + \epsilon_i \label{toy-model}
\end{align}
where Assumptions \ref{strong_monotonicity} (strong monotonicity) and \ref{er} (exclusion restriction) hold and $\epsilon_i \sim N(0, \sigma^2)$. Because the exclusion restriction holds, $E[\thetahat\stsls] = E[\thetahat\siv] = \theta$. However, due to violation of the NCE assumption, the principal-score and as-treated estimators (for example) will be biased. 

First, consider a comparison of $\thetahat\siv$ and $\thetahat\sps$. It is straightforward to show that $\E[\thetahat\sps] = \theta + (1-\pihat_c)\alpha_c$. The degree of bias for the CACE incurred by the principal-score estimator depends on the compliance proportion $\pihat_c$ and the compliance effect $\alpha_c$. While the principal-score estimator incurs this bias, it is more efficient than the IV estimator. Following the argument in \cite{feller2017principal}, one can show that the variance of $\thetahat\sps$ is
\begin{align}
\var(\thetahat\sps | \bX, \bS, \bC) &= \var\left(\left.n_{11}\inv\sum_{i : S_i = 1, Z_i = 1} Y_i\right|\bX, \bS, \bC\right) + \var\left(\left.n_0\inv\sum_{i : Z_i = 0} \frac{Y_i\ehat(\bX_i)}{\pihat_c}\right| \bX, \bS, \bC\right)\\
&= \sigma^2\left(\frac{1}{n_{11}} + \frac{n_0\inv\sum_{i : Z_i = 0}\ehat(\bX_i)^2}{n_0\pihat_c^2}\right)
\end{align}
On the other hand, the variance of $\thetahat\siv$ can be shown to be
\begin{align}
\var(\thetahat\siv | \bX, \bS, \bC) &= \var\left(\left.\frac{n_1\inv\sum_{i: Z_i = 1}Y_i}{\pihat_c}\right|\bX, \bS, \bC\right) + \var\left(\left.\frac{ n_0\inv\sum_{i: Z_i = 0}Y_i}{\pihat_c}\right| \bX, \bS, \bC\right)\\
&= \sigma^2\left(\frac{1}{n_{11}}\times\frac{n_1}{n_{11}} + \frac{1}{n_0\pihat_c^2}\right)
\end{align}
Noting that $\ehat(\bX_i) \in (0,1)$ so $n_0\inv\sum_{i : Z_i = 0}\ehat(\bX_i)^2 < 1$ and $n_1 > n_{11}$, it is clear that 
\[
\var(\thetahat\siv | \bX, \bS, \bC) > \var(\thetahat\sps | \bX, \bS, \bC).
\]

Similarly, using a classic omitted-variable result, because the as-treated model is missing only the variable for compliance $C_i$, $E[\thetahat\sat] = \theta + \alpha_c\eta$ where $\eta$ is the coefficient for $S_i$ in the least squares regression of $C_i$ onto $S_i$ and $\bX_i$. The quantity $\eta$ will also be $1 - \pi_c$ if $\bX_i \amalg S_i$. Comparing $\thetahat\sat$ to $\thetahat\stsls$ again shows the bias-variance tradeoff in the candidate estimators. The TSLS estimator and the as-treated estimator arise from similar models, but $\thetahat\stsls$ uses the estimated $\Shat_i$ in place of $S_i$. Because of using the estimated quantity, the TSLS estiator will incur additional variability and \[
\var(\thetahat\stsls | \bX, \bS, \bC) \geq \var(\thetahat\sat | \bX, \bS, \bC)
\] (cf. the result in \cite{murphy2002estimation}). 

Our approach seeks to exploit this bias-variance tradeoff, borrowing information from the possibly biased estimators (like $\thetahat\sat$ and $\thetahat\sps$) to induce greater efficiency in the SCE.

\section{Synthetic estimation}

\subsection{The estimator}
In this section, we propose a class of SCEs which leverages the information in all candidate estimators. 
Let the set of candidate estimators be denoted as $\bthetahat = (\thetahat_0, \thetahat_1, ..., \thetahat_k)' = (\thetahat_0, \bthetahat_1')'$, where $\thetahat_0$ is an estimator that can be presumed to be unbiased, and $\bthetahat_1$, a vector of length $k$, collects all other candidates.  Because the exclusion restriction (Assumption \ref{er}) can often be plausibly assumed to hold, we typically consider either the IV estimator or the TSLS estimator to be $\thetahat_0$.  When Assumption \ref{er} is deemed unlikely, another estimator may be considered as $\thetahat_0$. We consider synthetic estimators as a convex combination of the candidates

\begin{equation} \label{synthest}
\theta_s(\bthetahat , \bb_1) = (1 - \mathbf 1' \mathbf b_1) \thetahat_0 + \bb_1'\bthetahat_1
\end{equation}
where all entries of $\bb_1$ are between 0 and 1, and $\mathbf 1' \mathbf b_1 \leq 1$. The synthetic estimator in \eqref{synthest} is written as a function of $\bthetahat$ and $\bb_1$ to illustrate that different synthetic estimators are possible with different candidate estimators $\bthetahat$ and different weight vectors $\bb_1$. 

The synthetic estimator aims to lower the MSE of the supposedly unbiased $\thetahat_0$ by including the possibly biased $\bthetahat_1$ in the hopes of attaining lower variance without incurring too much estimation bias.  To achieve this goal, the convex combination would ideally be chosen to directly minimize the MSE.  We adopt the same rationale as in \citep{robinson1991blup, longford2006missing} to derive the SCE. Specifically, 
let the sampling variance of the candidate estimators be
\begin{equation*}
\bSigma_n = \var \begin{pmatrix} \hat\theta_0 \\ \boldsymbol {\hat\theta}_1  \end{pmatrix} =
\begin{pmatrix}     V_{0,n} & \mathbf C'_n\\
                             \mathbf C_n     & \mathbf V_n \end{pmatrix},
\end{equation*}
where the small $n$ in the subscript denotes the finite sample size. 
Let the biases of candidate estimators $\bthetahat_1$ be denoted as
\begin{equation*}\begin{split}
    &\mathbf d_n = E(\bthetahat_1 - \theta)
\end{split}\end{equation*}
Given $\bb_1$, the MSE of \eqref{synthest} is
\begin{equation}\begin{split}\label{bMSE}
 \quad E\left[\left\{ \theta_s(\bthetahat , \bb_1) - \theta\right\}^2\right] 
	&= (1 - \mathbf 1' \mathbf b_1)^2   V_{0,n} + \mathbf {b}_1' \mathbf{ V}_n \mathbf b_1  + 2 (1-\mathbf{1'b}_1 )\mathbf{ C}'_n  \mathbf{b}_1 + (\mathbf{b}_1'{\mathbf{d}}_n)^2  \\
    &=  V_{0, n} - 2{\mathbf P}_n' \mathbf b_1 + \mathbf b_1'({\mathbf T}_n + {\mathbf D}_n )\mathbf b_1,
\end{split}\end{equation}
where 
\begin{displaymath}\begin{split}
&{\mathbf P}_n = V_{0, n} - {\mathbf{ C}}_n, \\
&{\mathbf T}_n =  V_{0,n} - \mathbf{1  C'}_n - \mathbf{ C}_n \mathbf{1'} + \mathbf{ V}_n, \\
&{\mathbf D}_n = {\mathbf{d}}_n {\mathbf{d}}_n' ,
\end{split}\end{displaymath}

Let $b_1^*(\bSigma_n , \bd_n)$ be the minimizer of \eqref{bMSE}, which is expressed as a function of $\bSigma_n$ and $\bd_n$ to remind readers that it is a function of the bias and sampling variance of the candidate estimators. Suppose $b_1^*(\bSigma_n , \bd_n)$ is on the interior of the convex constraint. Then, 
\begin{equation} \label{eqn:opt_known}
b_1^*(\bSigma_n , \bd_n) = (\mathbf T_n + \mathbf D_n)^{-1}\mathbf P_n .
\end{equation}
This solution assumes that $\bSigma_n$ and $\bd_n$ are known. In practice, they are unknown and can be replaced with estimators. The minimizer of \eqref{bMSE} with plug-in estimates for the sampling moments is
\begin{equation} \label{eqn:opt}
b_1^*(\widehat{\bSigma}_n , \widehat{\bd}_n) = (\widehat{\mathbf T}_n + \widehat{\mathbf D}_n)^{-1}\widehat{\mathbf P}_n .
\end{equation}
If $b_1^*(\widehat{\bSigma}_n , \widehat{\bd}_n)$ is outside of the boundaries of the convex constraint, then \eqref{eqn:opt} needs to be projected to the boundaries of the constraints. The general form of our proposed synthetic estimator plugs the estimated optimal weight from \eqref{eqn:opt} into \eqref{synthest}

\begin{equation}
\thetahat_s = \theta_s\left\{\bthetahat~,~ b_1^*(\widehat{\bSigma}_n , \widehat{\bd}_n)\right\}
\end{equation}
which is written as a function of $\bthetahat$, $\bSigmahat_n$, and $\bdhat_n$ to illustrate that different synthetic estimators are possible with different candidate estimators $\bthetahat$, different estimates of the sampling variance of the candidate estimators $\bSigmahat_n$, and different estimates of the bias of the candidate estimators $\bdhat_n$. 

\subsection{Bias estimation}
It is typically straightforward to compute $\bSigmahat_n$ (and thus $\widehat{\mathbf P}_n$ and $\widehat{\mathbf T}_n$) by using a resampling-based method, e.g., the nonparametric bootstrap, jackknife, or random grouping (see, e.g., \citep{kovar1988bootstrap}). In what follows, $\bSigmahat_n$ can be any of these estimates of $\bSigma_n$. By contrast, it is much more difficult to estimate the finite-sample biases of candidate estimators. We propose three ways for computing $\bdhat_n$.

\paragraph{Raw differences between candidate estimators and $\thetahat_0$.}
Because we assume that $\thetahat_0$ is unbiased, we can compute a crude estimate of the bias in the other candidate estimators by taking their raw differences with $\thetahat_0$:
\[
\bDeltahat_n = (\thetahat_1 - \thetahat_0, ..., \thetahat_k - \thetahat_0)'.
\]
Denote the SCE using the raw difference as $\thetahat_{s-\text{raw}} = \theta_s\left\{\bthetahat~,~ b_1^*(\bSigmahat_n , \bDeltahat_n)\right\}$.

\paragraph{Shrinking raw differences.}
Unless two candidate estimators are highly correlated, their raw differences can be highly variable. To ameliorate the variability, we may want to regularize the bias estimates by down-weighting the most highly variable ones: 
\[
\bDeltahat_{\text{shrunk}} = \bw \circ \bDeltahat_n
\]
where $\circ$ denotes element-wise multiplication and $\bw = (w_1, ..., w_k)$ with
\[
w_j = \frac{ (\bDeltahat_n)_j^2}{\Vhat_{0,n} + (\bVhat_{n})_{j,j} - 2(\bChat_{n})_j + (\bDeltahat_n)_j^2}
\]. 

The expression for the weight is derived by choosing a shrinkage value that minimizes the mean squared error of the bias estimate, i.e. $w_j = \argmin w \text{MSE}( w(\thetahat_j-\thetahat_0) )$. If the bias $(\bDeltahat_n)_j$ is large relative to the variance $\widehat{\var}(\thetahat_1 - \thetahat_0) = \Vhat_{0,n} + (\bVhat_{n})_{j,j} - 2(\bChat_{n})_j$, then the weight will be close to 1, meaning the bias is not shrunk very much. On the other hand, if the bias is small compared to the variance, the weight $w_j$ will be close to 0 and the bias will be shrunk toward 0. This results in the estimator $\thetahat_{s-\text{shrunk}} = \theta_s\left\{\bthetahat~,~ b_1^*(\bSigmahat_n , \bDeltahat_{\text{shrunk}})\right\}$.

\paragraph{Sample-splitting approach.}
We also consider an approach which estimates the bias on an independent subset of data. Consider splitting the available data into two equally sized datasets, and estimating the candidate estimators on each. Let $\bthetahat_{\Asc}$ be the set of candidates estimated on one half, and $\bthetahat_{\Bsc}$ be estimated on the other half. Similarly define $\bDeltahat_{\Asc}$ and $\bDeltahat_{\Bsc}$. We can estimate the optimal weights from \eqref{eqn:opt} in the two subsets of the data, but apply the weights computed on one half to the candidate estimators from the other half and average the two:

\begin{equation*}
\thetahat_{s-\text{split}} = \frac{1}{2} \theta_s\left\{\bthetahat_{\Asc}~,~ b_1^*(\bSigmahat_n , \bDeltahat_{\Bsc})\right\}  + \frac{1}{2} \theta_s\left\{\bthetahat_{\Bsc}~,~ b_1^*(\bSigmahat_n , \bDeltahat_{\Asc})\right\}
\end{equation*}

Note that in the sample-splitting estimator, a single estimate of $\bSigma_n$ is used in estimating the weights on both halves of the data. This single estimate $\bSigmahat_n$ is estimated using the full sample. We take this approach for its computational simplicity and because the variability in estimation of $\Sigma_n$ is typically of a lower order than the variability in estimation of $\bd_n$.

\section{Asymptotic behavior}
\subsection{Asymptotic distribution of synthetic estimator}
In this section, we will establish the asymptotic distribution of the proposed estimator using the raw differences as the bias estimate. We make the following assumptions on the asymptotic behavior of candidate estimators as well as their sampling variance and the corresponding estimators. 
\begin{assumption} \label{asymp-assump1} As $n \to \infty$, 
\begin{displaymath}
\sqrt{n}  \begin{pmatrix} \hat\theta_0 - \theta \\  \boldsymbol{\hat \Delta}_n \end{pmatrix} \xrightarrow{d} 
\begin{pmatrix} J_0 \\ \mathbf J \end{pmatrix} \sim
    N\Bigg( \begin{pmatrix} 0 \\ \mathbf d  \end{pmatrix},  \begin{pmatrix}    V_0 & -\mathbf P'  \\
                             -\mathbf P  & \mathbf T  \end{pmatrix} \Bigg).
\end{displaymath}
\end{assumption}

\begin{assumption}\label{asymp-assump2}
 As $n \to \infty$,
\begin{displaymath}   
n\bSigma_n = \begin{pmatrix}    n V_{0,n} & n\mathbf C'_n\\
                             n\mathbf C_n     & n\mathbf V_n \end{pmatrix}
\rightarrow 
\begin{pmatrix}     V_0 & \mathbf C'\\
                             \mathbf C     & \mathbf V \end{pmatrix}.
\end{displaymath}
\end{assumption}

\begin{assumption}\label{asymp-assump3}
\begin{displaymath}
\bSigmahat_n
= \bSigma_n + o_p(n^{-1}).
\end{displaymath}
\end{assumption}

Assumption \ref{asymp-assump1} states that all candidate estimators have the usual root-n convergence rate. Assumption \ref{asymp-assump2} further assumes that candidate estimators are essentially uniformly integrable so that the sampling moments are also converging.  By these two assumptions, $\hat \theta_0$ is unbiased asymptotically but  other candidate estimators may not be.  When a candidate estimator is inconsistent, i.e., an entry in $\mathbf d$ is infinite, the corresponding weight in $\hat{\mathbf b}$ converges to zero in probability.  Here, we only consider candidate estimators with a finite asymptotic bias, i.e., all entries of  $\mathbf d$ are finite.  Assumption \ref{asymp-assump3} is the standard rate of the sampling variance estimator in parametric models and the usual bootstrap variance estimator.

Let $\mathbf P = V_0-\mathbf C$ and $\mathbf T = V_0 + \mathbf V - \mathbf 1 \mathbf C' - \mathbf C \mathbf 1'$. Let  $K =  -\frac{\mathbf {P' T^{-1}JJ'T^{-1}J}}{1+\mathbf{J'T^{-1}J}}
 = -a \mathbf {P'T^{-1}J}$, where $a = \frac{\mathbf{J'T^{-1}J}}{1+\mathbf{J'T^{-1}J}}$ is a random variable with a support of $(0, 1)$. Further define two scalar constants $\rho, \lambda$ as
\begin{equation}\begin{split}\label{eqn:consts}
	&E(K) = -\rho \mathbf {P'T^{-1}d}. \\
	&E(K^2) = \lambda  \mathbf{P'T}^{-1} (\mathbf T + \mathbf{dd'}) \mathbf{T}^{-1}\mathbf{P}.
\end{split}\end{equation}

~\\

\begin{flushleft}
\textbf{Lemma 1.} Some useful facts of the random variable $K$ and constants $\rho, \lambda$:
\end{flushleft}
\begin{enumerate}
    \item  $\var(K) =   \lambda \mathbf{P'T}^{-1} (\mathbf T + \mathbf{dd'}) \mathbf{P'T}^{-1}  - \rho^2 (\mathbf{P'T^{-1}d})^2;$
    \item $0 \le \lambda \le 1$
\end{enumerate}

~\\

The following theorem gives the asymptotic behavior of the synthetic estimator. 

~\\

\begin{flushleft}
\textbf{Theorem 1.} By Assumptions \ref{asymp-assump1}, \ref{asymp-assump2}, and \ref{asymp-assump3}, as $n\to \infty$, 
\end{flushleft}
\begin{displaymath}
\sqrt{n} (\hat \theta_{s-\text{raw}} - \theta) 
\xrightarrow{d} G = J_0 + \mathbf P'  (\mathbf T +  \mathbf{J'J} )^{-1} \mathbf J, 
\end{displaymath}
where 
\begin{align}
E(G^2) =  V_0 +    (\lambda-1)\mathbf{P'T}^{-1} \mathbf P + (\mathbf{P'T^{-1}d} )^2 (1-2\rho +\lambda).\label{th1}
\end{align}

~\\

\proof
By Assumptions \ref{asymp-assump2} and \ref{asymp-assump3}, $n\bPhat_n = \mathbf P + o_p(1)$, $n\bThat_n = \mathbf T + o_p(1)$. Then
\begin{equation}\begin{split}\label{eqn1}
 b_1^*(\bSigmahat_n , \bDeltahat_n) &=
 (\bThat_n + \bDeltahat_n\bDeltahat_n')^{-1}\bPhat_n 
 =(\mathbf T + n\bDeltahat_n\bDeltahat_n')^{-1}\mathbf P+ o_p(1) \\
\end{split}\end{equation}

By Slutsky's theorem, \eqref{eqn1}, and Assumption \ref{asymp-assump1}, as $n \to \infty$
\begin{displaymath}
\sqrt{n} (\hat \theta_{s-\text{raw}} - \theta) =  \sqrt{n} (\hat\theta_0 - \theta) + 
\sqrt{n}  b_1^*(\bSigmahat_n , \bDeltahat_n)' \boldsymbol{\hat \Delta} 
\xrightarrow{d} J_0 + 
\mathbf P' (\mathbf T + \mathbf{J'J})^{-1} \mathbf J = G. 
 \end{displaymath}

To derive $E(G^2)$, we use the relationship $E(G^2) = \var(G) + [E(G)]^2$.  By a basic theorem in linear algebra,
 \begin{displaymath}\begin{split}
  (\mathbf T + \mathbf{JJ'})^{-1}    & = \mathbf T^{-1} -\frac{\mathbf T^{-1}\mathbf{JJ'}\mathbf T^{-1}}{1+\mathbf{J'}\mathbf T^{-1}\mathbf{J}}.
\end{split}\end{displaymath}
Also note
$ 	J_0 | \mathbf J \sim N( -\mathbf P'\mathbf{T}^{-1}(\mathbf J - \mathbf d), 
 	                                                       V_0 - \mathbf P' \mathbf{T}^{-1} \mathbf P ).$ We then have
 \begin{displaymath}\begin{split}
&E(G ) = \mathbf{P'T^{-1}d} + E (K) = (1-\rho)\mathbf{P'T^{-1}d}, \\
 & E [ \var( G | \mathbf J) ] = V_0  - \mathbf{P'T^{-1}P},\\
 & \var [ E ( G |  \mathbf J) ]
   = \var \left ( -\frac{\mathbf P' \mathbf{T^{-1}JJ T^{-1}J}}{1+\mathbf{J'T^{-1}J}}  \right )
    = \var (K).
  \end{split}
 \end{displaymath}
Combine these terms together and by Lemma 1,
 \begin{displaymath}\begin{split}
\E[ G^2 ] &   
  =   (1-\rho)^2(\mathbf{P'T^{-1}d} )^2
      +   V_0  - \mathbf{P'T^{-1}P}  
      +\lambda \mathbf{P'T}^{-1} (\mathbf T + \mathbf{dd'}) \mathbf{P'T}^{-1}  - \rho^2 (\mathbf{P'T^{-1}d})^2\\
      &=V_0  + (\lambda-1)\mathbf{P'T}^{-1} \mathbf P
          +(\mathbf{P'T^{-1}d} )^2 [ (1-\rho)^2  + \lambda -\rho^2]\\
       &=V_0  +    (\lambda-1)\mathbf{P'T}^{-1} \mathbf P +
            (\mathbf{P'T^{-1}d} )^2 ( 1 -2\rho +\lambda).
 \end{split}\end{displaymath}
  $\square$

The second moment $E(G^2)$ represents the asymptotic efficiency measure of the synthetic estimator. Compared with the unbiased estimator $ \hat \theta_0$, the synthetic estimator has a better asymptotic efficiency if and only if 
\begin{equation}
(\lambda-1)\mathbf{P'T}^{-1} \mathbf P +
            (\mathbf{P'T^{-1}d} )^2 ( 1 -2\rho +\lambda) < 0.
\end{equation}
Since $0<\lambda<1$, a sufficient condition for efficiency gain by $\hat \theta_{s-\text{raw}}$ is
\begin{equation}
    1 -2\rho +\lambda < 0.  
\end{equation}
When all candidate estimators have zero asymptotic bias, that is $\mathbf d = \mathbf 0$, the synthetic estimator is guaranteed to be asymptotically more efficient than $\hat \theta_0$.   

\subsection{Inference}\label{inference}
We may use a plug-in version of $E[G^2]$, as found in \eqref{th1}, to compute confidence intervals. Because the SCE is based on estimators that may be biased, it is reasonable to anticipate at least a small amount of bias in the SCE. Therefore, using the mean squared error $E[G^2]$ may be preferred to $\var(G)$ for creating confidence intervals because it incorporates the bias and produces wider intervals.

A (1-$\alpha$)\% confidence interval can be constructed as
\[
\thetahat_s \pm z_{1-\alpha/2}\widehat{\varsigma}
\]
where $z_q$ is the $q$th quantile of the standard normal distribution and
\[
\widehat{\varsigma} = \Vhat_{0,n} +    (\lambdahat-1)\mathbf{\Phat_n'\That_n}^{-1} \mathbf \Phat_n + (\mathbf{\Phat_n'\That_n^{-1}\dhat_n} )^2 (1-2\rhohat +\lambdahat)
\]
where $\rhohat = \frac{\Ehat[K]}{\Phat_n'\That_n^{-1}\bdhat_n}, \lambdahat = \frac{\Ehat[K^2]}{ \mathbf{\Phat_n'\That_n}^{-1} (\mathbf \That_n + \mathbf{\bdhat_n\bdhat_n'}) \mathbf{\That_n}^{-1}\mathbf{\Phat_n}}$, $K = -\frac{\mathbf {\Phat_n' \That_n^{-1}JJ'\That_n^{-1}J}}{1+\mathbf{J'\That_n^{-1}J}}$ and $\bJ \sim N(\bdhat_n, \bThat_n)$. In practice, simulation may be used to generate a large collection of $\bJ$s from which the moments of $K$ may be empirically estimated. Let $\{\bJ_s\}_{s=1,...,\Ssc}$ be such a large collection of $\bJ$s drawn from $N(\bdhat_n, \bThat_n)$. Then, one may compute
\begin{align*}
    \Ehat[K] = \Ssc\inv\sum_{s=1}^\Ssc -\frac{\mathbf {\Phat_n' \That_n^{-1}J_sJ_s'\That_n^{-1}J_s}}{1+\mathbf{J_s'\That_n^{-1}J_s}}\\
    \Ehat[K^2] = \Ssc\inv\sum_{s=1}^\Ssc \left(\frac{\mathbf {\Phat_n' \That_n^{-1}J_sJ_s'\That_n^{-1}J_s}}{1+\mathbf{J_s'\That_n^{-1}J_s}}\right)^2
\end{align*}
and the confidence intervals follow.

\section{Simulation}

We performed Monte Carlo experiments to demonstrate the finite-sample performance of the proposed synthetic compliance estimators. These simulations demonstrate how compliance effects (in violation of Assumption \ref{nce}) and sample size impact the performance of the proposed SCEs. 

We consider two data-generating mechanisms. First, we adapt the simulation set-up in \cite{Ding2017}. In this setting, the data-generating model takes the following form
\begin{align}
    Y_i(1) = \sum_{j=1}^5 X_{ij} + 2C_i + 1 + \epsilon_i\\
    Y_i(0) = \sum_{j=1}^5 X_{ij} + 2 + \epsilon_i
\end{align}
where $C_i \sim \text{Bernoulli}\{\pi_c(\bX_i)\}$ is an indicator of being a complier, $X_{ij} \sim N(0,1), j = 1, ..., 4$, and $X_{i5} \sim \text{Bernoulli}\{0.5\}$ and $\logit\{\pi_c(\bX_i)\} = 0.5X_{i1} + 0.5X_{i2} + X_{i3} + X_{i4} + \eta X_{i5}$.
However, when the data are analyzed, $X_{i5}$ is omitted from the models, and therefore $\eta$ corresponds to a measure of the violation of Assumption \ref{nce}. When $\eta$ is large in magnitude, we expect more bias in the estimators that rely on the assumption of no compliance effect, such as $\thetahat\sat, \thetahat\spp$, and $\thetahat\sps$. We tested three different sample sizes $n = 200, 500, 1000$, and we let $\eta$ vary between -2 and 2.

The second data-generating mechanism is adapted from \citep{stuart2015assessing}. Here, the model has the following form:
\begin{align}
Y_i = \alpha_c C_i + \gamma_cC_iZ_i + \lambda_n X_i + (\lambda_c - \lambda_n)C_iX_i + \epsilon_i
\end{align}
where $C_i \sim \text{Bernoulli}\{\pi_c(X_i)\}$, with $\logit(\pi_c(X_i)) = \beta_0 + \beta_1X_i$, and $X_i \sim N(0,1)$. The CACE is identified by the parameter $\gamma_c$. The parameters $\alpha_c$ and $\lambda_c$ control the degree to which Assumption \ref{nce} -- no compliance effect -- is violated. If $\alpha_c > 0$, then compliers naturally have higher means than never-takers, and if $\lambda_c \neq \lambda_n$, then the effect of $X_i$ is different between compliers and never-takers. Both models ensure the exclusion restriction (Assumption \ref{er}) and strong monotonicity (Assumption \ref{strong_monotonicity}) hold.

Similarly to the first data-generating process, we generated data at three sample sizes, $n = 200, 500, 1000$, and we varied the level of violation of Assumption \ref{nce} by letting $\alpha_c$ take values in (0, 0.1, 0.2, 0.3, 0.4, 0.5).  We also allowed $\lambda_n, \lambda_c$, and $\gamma_c$ to be 0 or 1. We compared the performance of a range of synthetic estimators to the performance of the candidate estimators. In all simulations, $\Sigmahat$ was estimated using 200 bootstrap samples. 

\subsection{Overall performance of the SCE}
\begin{figure}
\centering
\begin{tabular}{c}
\includegraphics[width =\textwidth]{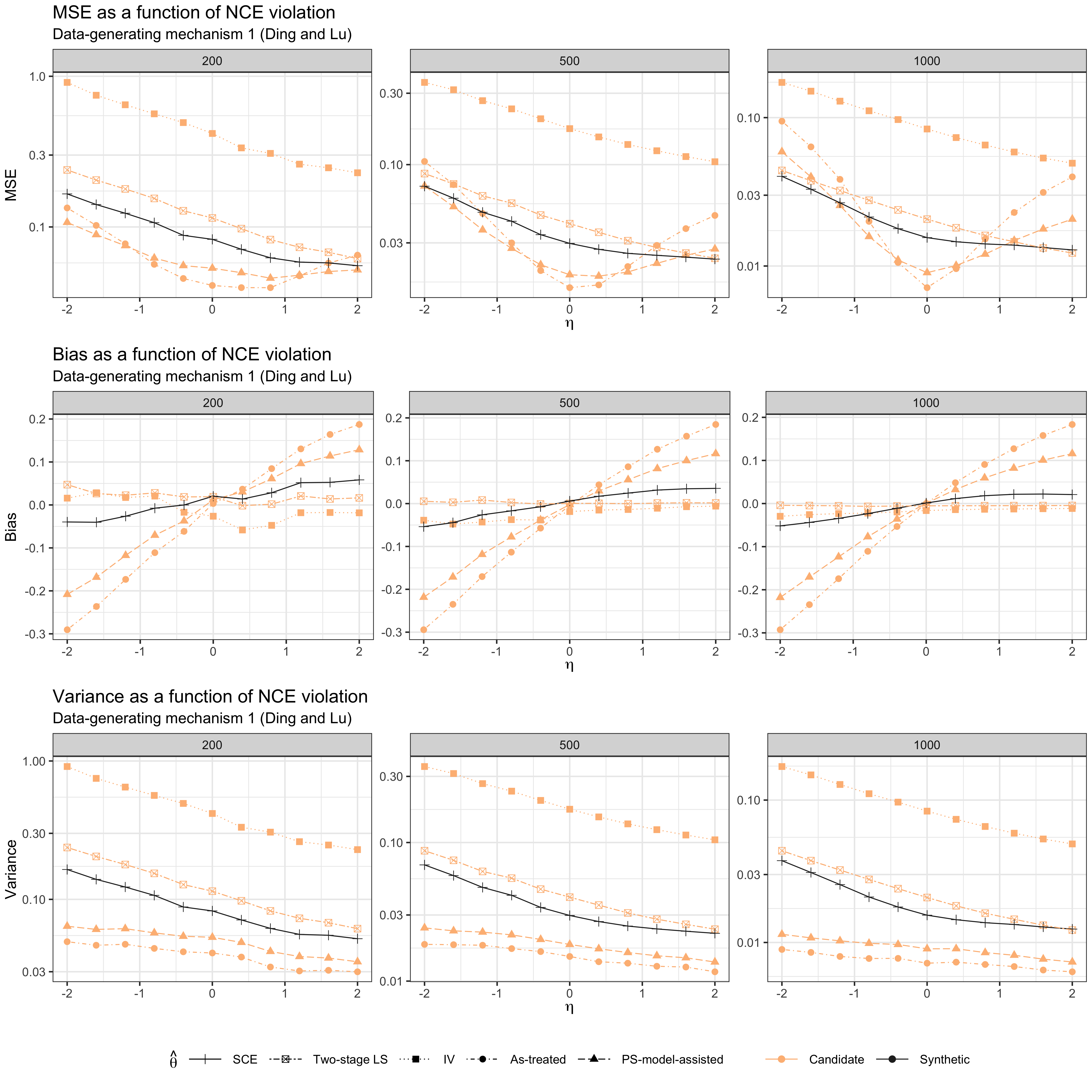}
\end{tabular}\vspace{0.2in}
\caption{MSE, bias, and variance of the SCE with $\thetahat_0 = \thetahat\stsls$ compared to four of the candidate estimators for the CACE as $\eta$ changes in data-generating mechanism 1. The SCE includes information from all candidate estimators, but only four candidates are shown here. Sample sizes ($n = 200, 500, 1000$) are given in different facets.The y-axis for MSE and variance are given on the log-scale.}
\label{sec-sim-pl}
\end{figure}

Figure \ref{sec-sim-pl} demonstrates the behavior of the SCE $\thetahat_{s-\text{raw}}$ with $\thetahat_0 = \thetahat\stsls$ compared to $\thetahat\stsls, \thetahat\siv, \thetahat\sat,$ and $\thetahat\saps$ in data-generating mechanism 1. We included all candidate estimators in the SCE, but we compare the behavior of the SCE to a few instructive candidate estimators, so as not to clutter up results with extraneous information.  It is clear from the figure that the SCE strikes a balance between the low bias and high variance of $\thetahat\stsls$ and the high bias and low variance of $\thetahat\sat$ and $\thetahat\saps$. It took advantage of the variance reduction inherent in combining multiple estimators without incurring too much bias. For every value of $\eta$ and at all sample sizes, the SCE had a variance and MSE lower than $\thetahat\stsls$. It was never as efficient as the low-variance candidate estimators like $\thetahat\sat$, but it also never incurred nearly as much bias as they did, either. While $\thetahat\sat$ had bias as high as 30\% of the CACE in magnitude, the SCE never had bias that was more than about 5\% of the CACE.

\begin{figure}
\centering
\begin{tabular}{c}
\includegraphics[width =\textwidth]{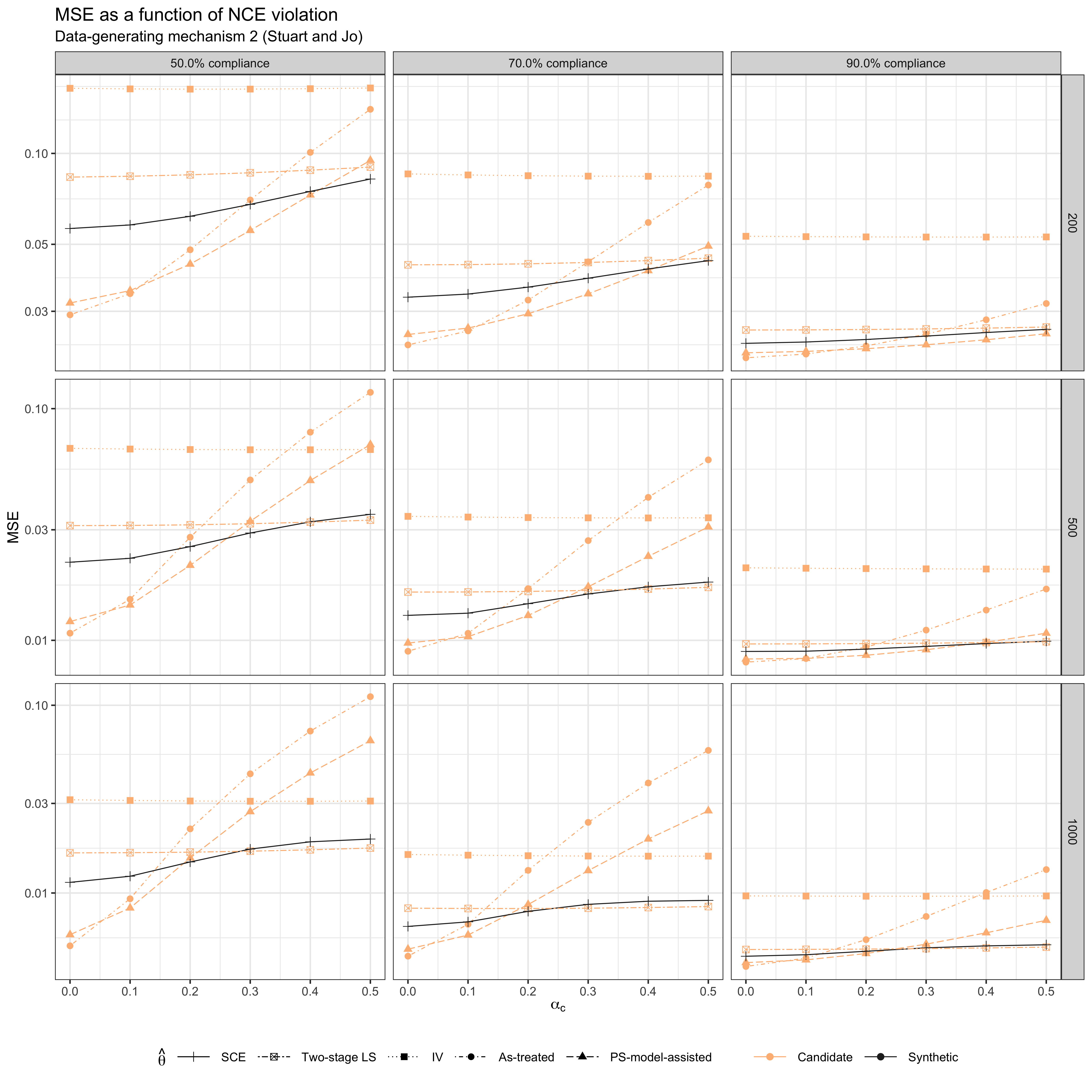}
\end{tabular}\vspace{0.2in}
\caption{MSE of SCE with $\thetahat_0 = \thetahat\stsls$ compared to four of the candidate estimators for the CACE as $\alpha_c$ changes in data-generating mechanism 2. The SCE includes information from all candidate estimators, but only four candidates are shown here. Sample sizes of $n = 200, 500, 1000$ are given in the rows and compliance rates are given in the columns. Simulation parameters are set to $\gamma_c = 0, \lambda_n = \lambda_c = 1, \beta_0 = 0.41, \beta_1 = 2$. The y-axis is shown on the log-scale to highlight differences.}
\label{main-mse-p}
\end{figure}

A similar robustness property was found in data-generating mechanism 2 (see Figure \ref{main-mse-p}. The SCE produces a sizable MSE reduction compare to $\thetahat\stsls$ when $\alpha_c$ is small and thus the bias and MSE of $\thetahat\sat$ and $\thetahat\saps$ are small. As the bias in $\thetahat\sat$ and $\thetahat\saps$ increase, the MSE of the synthetic estimator hovers at or near the MSE of $\thetahat\stsls$. The effects of creating a synthetic estimator are largest when compliance and sample size are both low (upper left panel in Figure \ref{main-mse-p}, where the SCE is the best estimator when $\alpha_c = 0.5$). 

\subsection{Effect of different bias estimates}
\begin{figure}
\centering
\begin{tabular}{c}
\includegraphics[width =\textwidth]{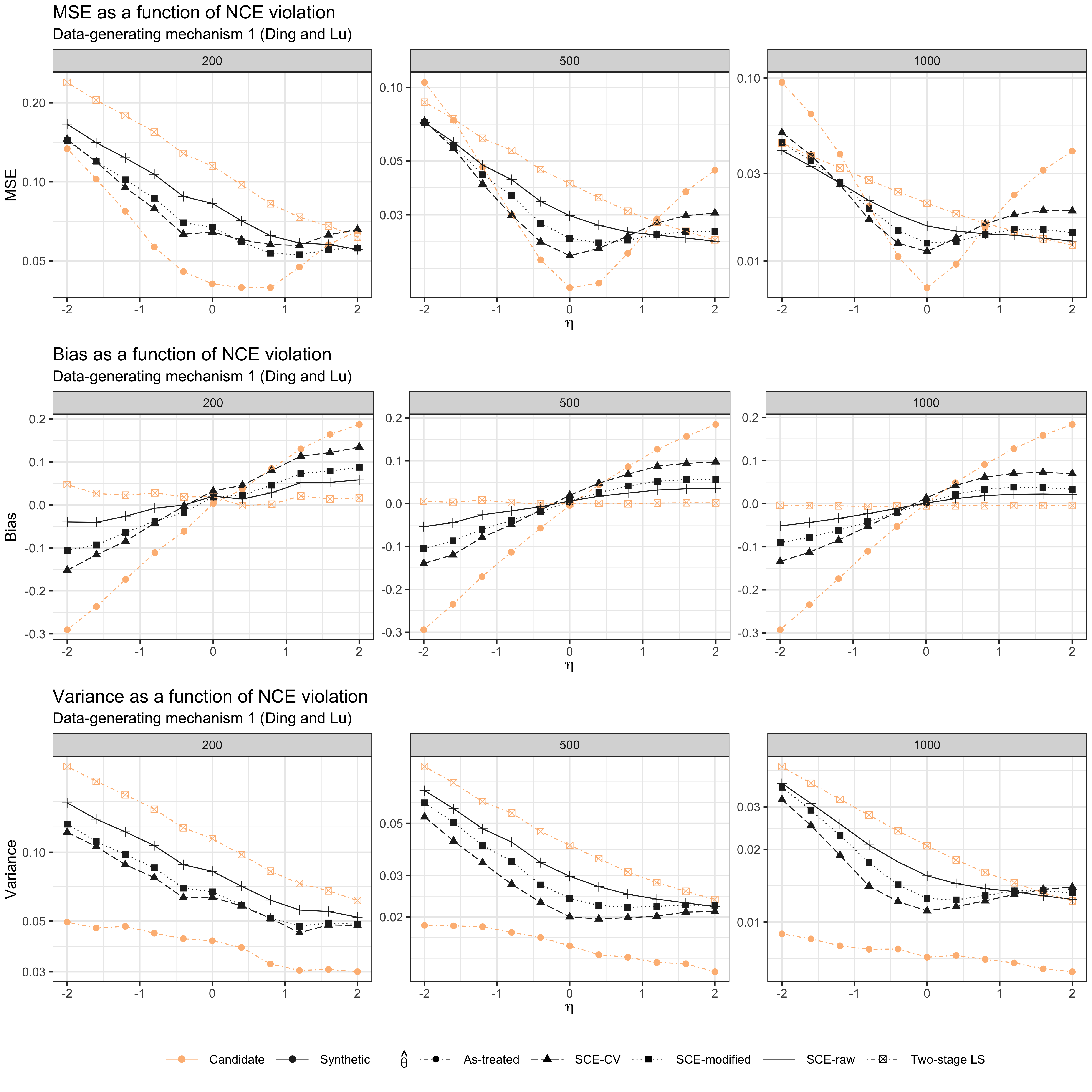}
\end{tabular}\vspace{0.2in}
\caption{MSE, bias, and variance of versions of the SCE with $\thetahat_0 = \thetahat\stsls$ and different bias estimates ($\bdhat_n$) as $\eta$ changes in data-generating mechanism 1. These are compared to $\thetahat\stsls$ and $\thetahat\sat$. The SCE includes information from all candidate estimators. Sample sizes ($n = 200, 500, 1000$) are given in different facets.The y-axis for MSE and variance are given on the log-scale.}\label{sec-sim-pl2}
\end{figure}

The performance of the SCE when using different estimates of the bias are shown in Figure \ref{sec-sim-pl2}. Three SCEs are shown, $\thraw, \thshr$, and $\thspl$, each with $\thetahat_0 = \thetahat\stsls$. These are compared to $\thetahat\stsls$ and the highest-bias, lowest-variance estimator, $\thetahat\sat$, which illustrates the relevant features. 

In the figure, $\thraw$ has performance closest to $\thetahat\stsls$ in terms of bias, variance, and MSE. As shown in Figure \ref{sec-sim-pl} as well, $\thraw$ took advantage of the other candidate estimators to decrease the variance of the TSLS estimator, while not incurring too much bias, even when the other estimators had a lot of bias. Using the alternative bias estimates loosened the relationship between $\thetahat_s$ and $\thetahat_0$, such that the SCEs took more advantage of the reduction in variance available from the other estimators, but at the same time suffered from higher bias when those estimators were more biased. In nearly all cases, $\thspl$ had the highest bias and lowest variance among the SCEs, with $\thshr$ falling between $\thspl$ and $\thraw$ in terms of both bias and variance. While $\thspl$ and $\thshr$ had markedly better performance than $\thraw$ when $\eta$ was close to 0 and all candidate estimators were unbiased, they also incurred upwards of 10-20\% bias in the extremes, and they even had worse performance than $\thetahat\stsls$ when sample size and $\eta$ were large ($n = 1000, \eta > 1$). 

\subsection{Inference on SCEs}
\begin{figure}
\centering
\begin{tabular}{c}
\includegraphics[width =\textwidth]{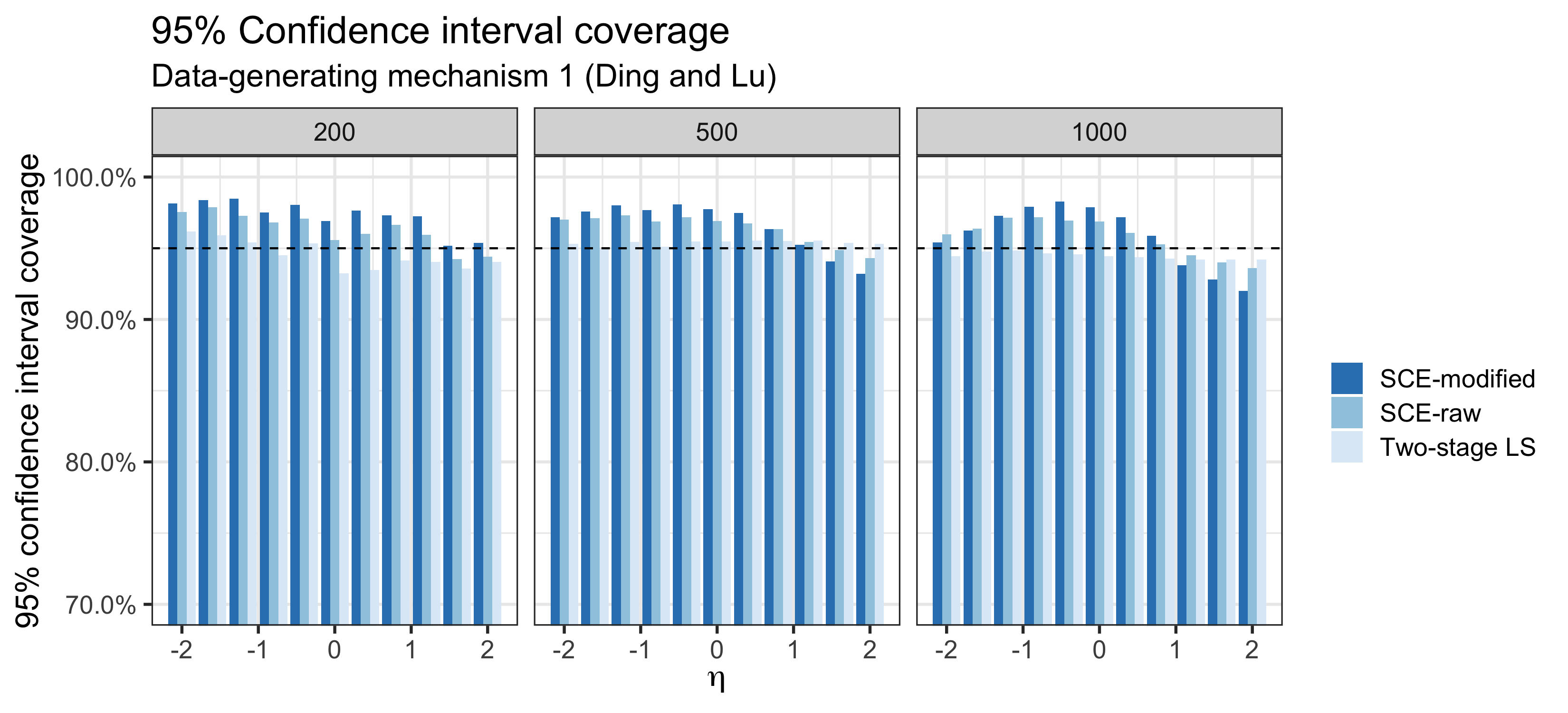}
\end{tabular}\vspace{0.2in}
\caption{95\% confidence interval coverage for $\thraw$, $\thshr$, and $\thetahat\stsls$ as $\eta$ changes in data-generating mechanism 1. Sample sizes ($n = 200, 500, 1000$) are given in different facets.}\label{ci-plot}
\end{figure}

The plug-in estimator of $E[G^2]$ given in Section \ref{inference} displays very good properties at all sample sizes. Figure \ref{ci-plot} shows 95\% confidence interval coverage for $\thraw$, $\thshr$, and $\thetahat\stsls$. The confidence interval coverage for $\thraw$ compares favorably to $\thetahat\stsls$, though confidence intervals based on $\Ehat[G^2]$ showed some over-coverage (as high as 98\%), while the standard bootstrap-based confidence intervals for $\thetahat\stsls$ had no such issues (coverage between 93\% and 96\%). Standard errors (SEs) for the SCEs were similarly slightly larger than their empirical standard errors, though this attenuates with sample size: the estimate SE for $\thraw$ was 0.04 too large when $n = 200$, 0.02 too large when $n = 500$, and 0.01 too large when $n = 1000$. 

\section{Conclusion}
While it is rather standard to use the IV approach to estimate the CACE in the presence of noncompliance, the IV approach usually suffers from a higher sampling variance compared with other alternative approaches. We have proposed a set of synthetic compliance estimators which maintains the low-bias properties of the IV and the TSLS estimators, while gaining efficiency from other biased estimators which usually have smaller sampling variance. The synthetic estimators are based on the classic rationale of balancing bias and variance. We derived asymptotic properties of one specific type of synthetic estimator and demonstrated the finite-sample properties by numeric simulations. The numerical study also showed that 95\% confidence intervals based on the asymptotic properties attained the nominal level of coverage, approximately. Despite the fact that the optimal synthetic estimator is always biased for the CACE, except in rare cases, we did not observe sizable biases of the proposed SCE in our simulation studies, particularly when employing raw differences, as in our asymptotic results. 

The main technical challenge in implementing the SCE is to estimate the biases of the possibly-biased candidate estimators. We proposed several approaches (raw difference, shrinking raw difference, and split sampling). In numerical studies we found that these three approaches differ in how they balance bias and variance.  More investigations are needed to better understand the phenomenon and to search for potentially better approaches to assess the sampling biases.

Our approach here could easily be generalized to include a larger class of estimators or to target a different estimand. Similar synthetic estimators could be constructed in any setting where many candidate estimators are available and the estimator that is presumed to be unbiased has a higher variance than the other candidates.

\bibliography{compliance_refs}
\bibliographystyle{plainnat}

\end{document}